\documentclass[preprint,12pt,3p]{elsarticle}

\usepackage{amssymb}
\usepackage{amsmath}
\usepackage{amsfonts,amsthm}
\usepackage{mathtools}
\usepackage{algorithmic}
\usepackage{algorithm}
\usepackage{array}
\usepackage{graphicx}
\usepackage{caption} 
\usepackage{booktabs} 
\usepackage{longtable}
\usepackage{multirow} 
\usepackage{array} 
\usepackage{subcaption}
\usepackage{float}
\usepackage{hyperref}
\usepackage{soul, xcolor} 
\usepackage{tabularx}
\usepackage{xcolor}
\biboptions{sort&compress}
\definecolor{myyellow}{RGB}{255, 255, 0}
\sethlcolor{myyellow} 
\graphicspath{{./figure/}}
\DeclareGraphicsExtensions{.pdf,.jpeg,.png,.jpg}

\journal{Energy Conversion and Management}

\begin{document}

\begin{frontmatter}



\title {Black-Start Power Capacity Sizing and Control Strategy for an Islanded DFIG Wind-to-Hydrogen System}
\tnotetext[1]{This document is the results of the research project funded by the China Datang Technology Innovation Co., Ltd. (10002552D24KJZB00009).}

\author[add1,add2]{Bosen Yang} 
\author[add1]{Kang Ma} 
\author[add2]{Jin Lin\corref{cor1}} 
\author[add2,add3]{Yonghua Song}
\cormark[1]
\ead{linjin@tsinghua.edu.cn}

\affiliation[add1]{organization={China Datang Technology Innovation Co.,Ltd.},
	city={Xiong’an},
	postcode={071800}, 
	country={China}}

\affiliation[add2]{organization={Department of Electrical Engineering},
            addressline={Tsinghua University}, 
            city={Beijing},
            postcode={100084}, 
            country={China}}

\affiliation[add3]
{organization={Department of Electrical and Computer Engineer-
ing},
            addressline={University of Macau}, 
            city={Macau},
            postcode={999078}, 
            country={China}}
  \cortext[cor1]{Corresponding author}
\begin{abstract}
This paper proposes a black-start method for an off-grid wind-to-hydrogen (W2H) system comprising a wind farm based on Doubly-Fed Induction Generators (DFIGs), proton exchange membrane fuel cells (PEMFCs) serving as the black-start power source, and a hydrogen production industry. The PEMFC is installed within the hydrogen industry to facilitate direct access to hydrogen fuel. Based on the microgrid topology and black-start scheme, this study innovatively sizes the rated capacity of the PEMFC through power flow analysis. The capacity must be sufficient to charge passive components such as transmission lines and transformers, provide rotor excitation, and supply wind turbine (WT) and electrolyzer (ELZ) auxiliaries during startup. The proposed system integrates wind–hydrogen coordinated control (WHCC) and hydrogen–storage coordinated control (HSCC). Under maximum power point tracking (MPPT) of the WTs, the ELZ follows power fluctuations to absorb wind output, ensuring stable voltage and frequency. Fixed-frequency control applied to either the DFIG or PEMFC converters enables DFIGs to retain conventional grid-following (GFL) operation, reducing converter development costs. For both control modes, this paper establishes the black-start sequence and formulates a comprehensive coordinated control strategy for the entire system. The entire control system is validated through simulations in the Matlab/Simulink. Results confirm that the calculated PEMFC capacity supports reliable black-start while the black-start control strategy ensures smooth system self-startup. Furthermore, the coordinated control strategy maintains stable frequency and voltage under fluctuating wind power, demonstrating the practicality and robustness of the proposed approach.
\end{abstract}

\begin{graphicalabstract}
\end{graphicalabstract}

\begin{highlights}
\item Research highlight 1
\item Research highlight 2
\end{highlights}


\end{frontmatter}



\section{Introduction}
\label{sec1}
\subsection{Background and Motivation}
Green hydrogen is emerging as a pivotal clean energy carrier with significant potential for  deployment across carbon-intensive industrial sectors \cite{ref1}. In steel manufacturing, hydrogen-based direct reduction of iron ore offers a promising alternative to conventional coke-based processes \cite{ref2}. In the chemical industry, green hydrogen enables the catalytic synthesis of green ammonia and methanol, serving as essential feedstocks and sustainable fuels \cite{ref3,ref4}. In petroleum refining, hydrogen plays a critical role in hydrodesulfurization, improving fuel quality while reducing environmental impact \cite{ref5}. Consequently, renewable energy-based hydrogen production has been recognized as a critical pathway toward deep decarbonization \cite{ref6}. \par
Many countries, including China, Germany, the Netherlands, Denmark, and Norway, have launched large-scale renewable hydrogen projects. Most are initially designed as grid-connected systems, benefiting from grid support. For example, the 150 MW wind-solar-hydrogen integration project in Duolun, China, achieves an annual $\mathrm{CO_2}$ emission reduction of 129,300 tons \cite{ref7}. However, such grid-connected approaches face four critical challenges: (1) Regulatory restrictions on renewable hydrogen projects regarding grid connection and disconnection capacities \cite{ref8}; (2) Additional costs from grid capacity charges, demand charges, and power factor penalties, which increase the levelized cost of hydrogen (LCOH); (3) Ambiguities in defining “green fuel” when a large share of grid electricity is consumed \cite{ref9}. (4) Limited grid absorption capacity, as evidenced by congestion issues in the Netherlands \cite{ref10}. \par
In this context, off-grid renewable hydrogen production is emerging as a more sustainable and technically viable pathway to carbon reduction \cite{ref11}. Off-grid W2H systems can be categorized into pure off-grid and process off-grid configurations \cite{ref8}, depending on whether auxiliary equipment power is supplied by wind energy or partially supported by the grid. Pure off-grid systems require more sophisticated system capacity design and coordinated control strategies. Regardless of configuration, the fundamental challenge lies in the black-start process: without grid support, dedicated black-start power sources are required to initialize the system.\par
On the one hand, complex and variable wind conditions lead to frequent start-stop operations in off-grid wind-hydrogen systems, making reliable black-start sequencing and control strategies essential for safety, rapid response and optimized hydrogen production. On the other hand, the sizing of the black-start power source has a direct impact on the LCOH, thereby influencing the economic competitiveness of off-grid hydrogen production.
\label{subsec1.1}
\subsection{Literature Review}
To date, the black-start process of islanded W2H systems has not been explicitly addressed in the literature. Existing studies instead focus on black-start strategies for wind farms, which provide valuable insights into power source sizing, startup sequences, and coordinated control.\par
\textit{(1) Black Start Power Source Capacity Sizing. }\par
\cite{ref12} optimized energy storage system (ESS) power and energy capacities by minimizing cost under constraints of turbine startup power, frequency stability and output smoothing. However, this study overlooked that the ESS only needs to energize one or a few WTs, as the remaining units can be energized by those already in operation. \cite{ref13} highlighted this aspect, showing that starting x WTs with ESS and then using them to energize the rest reduces the required black-start capacity to the minimum energy for x units, with overall requirements further determined by startup duration. Research on black-start capacity sizing remains limited, with most studies relying on simulations or experiments to derive quantitative constraints—such as switching, charging, and startup durations—that can be used for capacity verification. \cite{ref14} compared four grid-forming (GFM) strategies through simulation, analyzing active and reactive power demands during transformer energization and cable charging, and derived indicative capacity boundaries for ESS sizing. \cite{ref15} compared hard and soft charging strategies for ESS in offshore wind farms, quantifying differences in inrush current and transient power to delineate capacity margins. \cite{ref16} reported field test data including power/energy curves and switching sequences, providing benchmarks for capacity and overload verification. \cite{ref17} simulated active/reactive power trajectories of VSM-type GFM inverters during black-start, offering insights into capacity selection. \cite{ref18} futher assessed power–energy constraints and control boundaries of ESS in black-start and islanded operation, facilitating practical capacity validation.\par
\textit{(2) Black-Start Sequences and Control Strategies of Wind Farms} \par
Several studies explored black-start control strategies in wind farms. \cite{ref19,ref20,ref21} proposed predictive models based on voltage sensitivity and dynamic reactive power response, employing model predictive control (MPC) for coordinated active and reactive power regulation among wind farms, ESS, and Static Var Generators (SVGs). \cite{ref22} demonstrated that a grid-forming ESS can black-start an offshore wind farm by charging passive components, energizing turbines, supporting grid stability, and eventually synchronizing with other islands. \cite{ref23} proposed a four-stage offshore wind black-start strategy using an auxiliary step-down transformer to interface with the onshore MMC, ensuring reliable startup. \cite{ref24} developed a two-layer control framework: a Virtual Synchronous Generator (VSG)-based primary layer providing frequency and voltage support, and an MPC-based secondary layer allocating WT and ESS power to balance frequency recovery and rotor stability. \cite{ref25} introduced a decentralized droop-based control scheme enabling multiple turbines to share power according to local wind conditions, thereby balancing generation and load during black-start. Finally, \cite{ref26} validated the black-start capability by integrating a small storage device into the DC bus for excitation, enabling grid-forming control of voltage and frequency.

\label{subsec1.2}
\subsection{Contributions}
The main contributions of this paper are summarized as follows:\par
(1) This study deploys PEMFCs as black-start power sources at hydrogen production industry. By utilizing shared hydrogen storage tanks, it mitigates concerns related to hydrogen fuel availability, storage, and transport. This study establishes a systematic methodology for black-start power source capacity calculation, where for PEMFC, only power capacity needs to be considered, eliminating the need for energy capacity considerations. The calculation accounts for (i) auxiliary power consumption of wind farms and hydrogen production industry, and (ii) transmission losses obtained from power flow analysis. \par
(2) An innovative coordinated control strategy is proposed for off-grid W2H systems. In the HSCC scheme, the DFIG operates under MPPT, the ELZ acts as an active power slack node to balance microgrid active power through voltage regulation, and the PEMFC serves as a reactive power slack node while also maintaining system frequency. Within the HSCC framework, the DFIG operates as a PQ node, while ELZ and PEMFC are characterized as QV and P$\theta$ nodes respectively, representing novel node types that do not exist in either conventional power systems or grid-forming converter-based systems.\par
(3) Complete black-start protocols are developed for both WHCC \cite{ref27} and the newly proposed HSCC strategy. Their effectiveness is validated through detailed case studies in Matlab/Simulink, demonstrating reliable black-start capability, stable system operation, and robustness under fluctuating wind conditions.\par
The remaining structure of this paper is organized as follows: Section II describes an integrated approach for black-start power source capacity sizing, utilizing estimation methods in conjunction with power flow computational analysis. Section III presents two coordinated control strategies, WHCC and HSCC, for off-grid W2H systems during black-start operations, along with comprehensive black-start operational sequencing protocols. Section IV validates the proposed approach through case studies conducted in Matlab/Simulink. Finally, the conclusions are summarized in Section V.
\label{subsec1.3}
\section{Method of Black-Start Power Capacity Sizing}
\label{sec2}
\begin{figure}[t]
  \centering
  \includegraphics[width=5in]{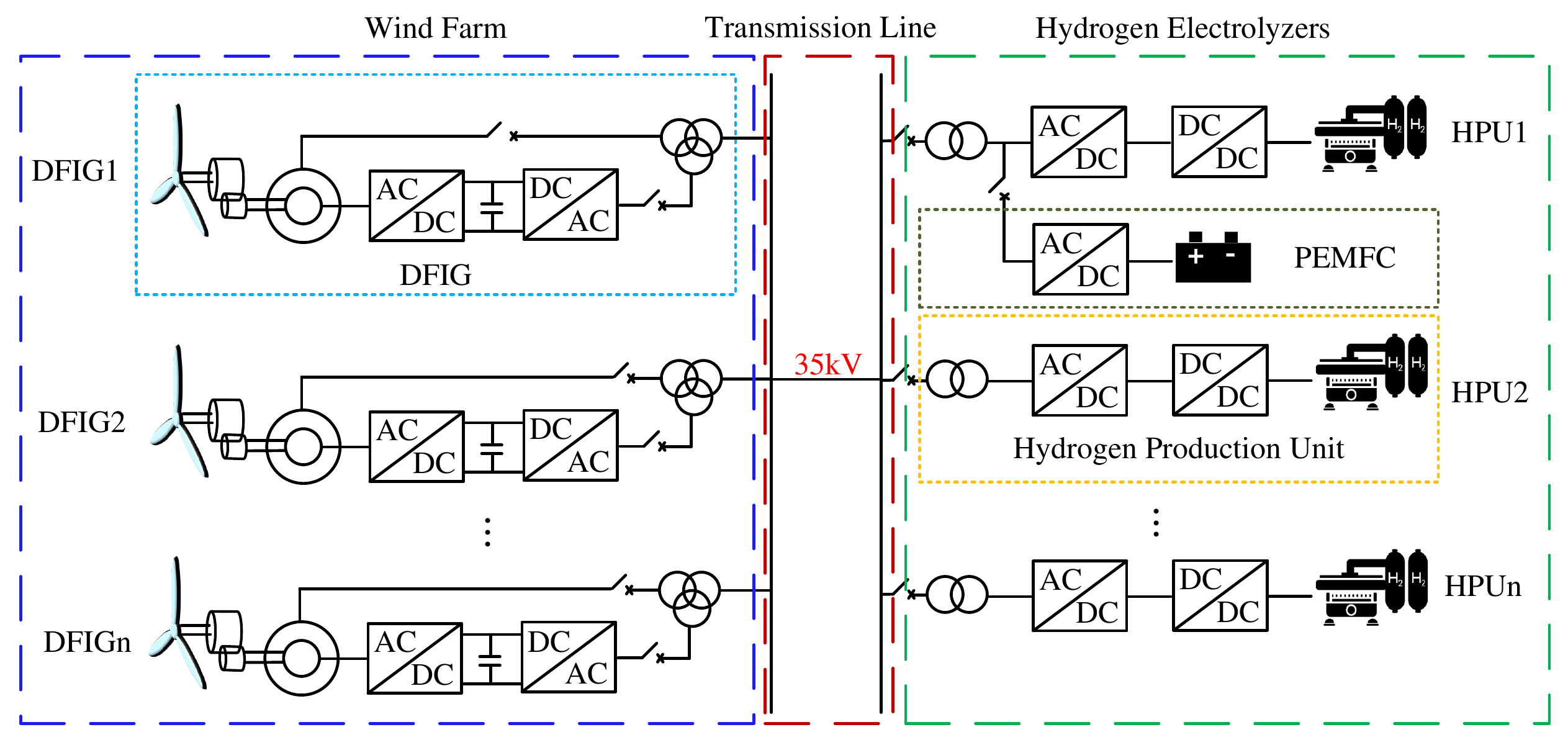}
  \caption{The topology of the islanded DFIG W2H system}
  \label{fig1}
\end{figure}
Fig.\ref{fig1} illustrates the overall topology of an islanded W2H system, in which a PEM fuel cell serves as the dedicated black-start power source. The system is composed of three main components: a wind farm, a transmission network and a cluster of hydrogen electrolyzers. Multiple DFIGs are connected in parallel through collector cables, while a 35kV transmission line is employed to deliver wind power at the 10MW scale over distances of up to 10km. In the hydrogen industry, PEMFCs are generally colocated with hydrogen production units, as direct hydrogen sourcing from the plant is more convenient than integration at the wind farm. \par
\begin{figure}[h]
  \centering
  \includegraphics[width=5in]{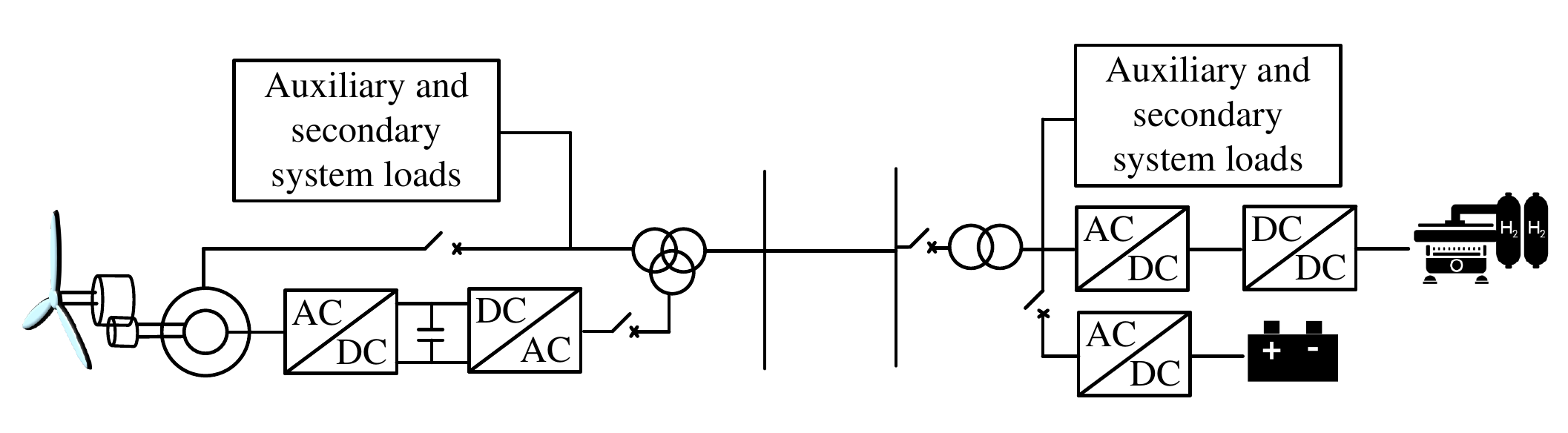}
  \caption{Schematic of the minimum power supply circuit driven by the PEM fuel cell during the black-start process}
  \label{fig2}
\end{figure}

Fig.\ref{fig2} illustrates that to ensure reliable black-start, the PEMFC capacity must be sufficient to supply (i) The minimum auxiliary and secondary loads of one DFIG and one hydrogen production unit, and (ii) the transmission losses incurred during the startup process. Accordingly, the following subsections detail the estimation of auxiliary loads and transmission losses, which together determine the required PEMFC rating.\par

\subsection{The minimum auxiliary and secondary system loads}
\label{subsec2.1}
At present, there is no standardized method for accurately modelling the minimum auxiliary loads of a DFIG or a hydrogen production unit. In this study, auxiliary load requirements during black-start are derived from operational data collected at an existing wind farm and a large-scale hydrogen production industry.\par
\begin{center}
\begin{table}[h]
  \centering
  \caption{Auxiliary Load of 1.5MW wind turbines}
  \label{tab1}
  \begin{tabular}{llll}
  \hline
  \textbf{}                     Electrical Devices                  &  1\#    & 2\#   & 3\#  \\ \hline
  Yawing motor(kW)                                & 12.0   & 8.8   & 12.0 \\
  Pitch drive motor(kW)                           & 12.0   & 8.4   & 12.0 \\
  Gearbox temperature control system(kW)          & 6.2    & 5.5   & 5.6  \\
  Gearbox lubrication system(kW)                  & 10.3   & 10.4  & 10.6 \\
  Generator temperature control system(kW)        & 15.4   & 1.1   & 11.0 \\
  Inverter temperature control system(kW)         & 0.0    & 0.8   & 0.8  \\
  Nacelle \& tower temperature control system(kW) & 22.7   & 26.5  & 23   \\
  Total(kW)                                       & 78.6   & 61.5  & 75   \\
  Ratio                                           & 5.24\% & 4.1\% & 5\%  \\ \hline
  \end{tabular}
  \end{table}
\end{center}
\begin{center}
\begin{table}[h]
\caption{Minimum Auxiliary Load for Black Start of Hydrogen Production Side}
\label{tab2}
\centering
\begin{tabular}{lcc}
\hline
Electrical Devices                & Number & Power (kW) \\ \hline
Alkali Circulation Pump             & 1      & 65        \\
Deaeration Electric Heater          & 1      & 100       \\
Hydrogen Dryer Electric Heater      & 1      & 120       \\
Alkali Preparation Pump             & 1      & 5.5       \\
Make-up Water Pump                  & 1      & 15        \\
Hydrogen Generation Control Cabinet & 1      & 2         \\
SIS Control Cabinet                 & 1      & 2         \\
Lighting Box                        & 1      & 5         \\
Comprehensive Building              & 1      & 500       \\
Central Control Room                & 1      & 500       \\
Field Cabinet Room                  & 1      & 200       \\
Total                               &        & 1634.5    \\ \hline
\end{tabular}
\end{table}
\end{center}

  Tbale.\ref{tab1} summarizes the auxiliary load of 1.5 MW DFIG WTs from three suppliers. It is noted that these loads vary with operating conditions such as wind speed and nacelle temperature. For example, when the nacelle temperature is within its normal range, the cooling system remains inactive, resulting in zero demand from that subsystem. To ensure sufficient design margin, this study assumes that all auxiliary devices operate at their rated power during black-start. Consequently, the minimum auxiliary load of the wind farm is approximated as 5\% of the rated power of a single DFIG. Once the first DFIG is energized, it can supply the auxiliary startup power for subsequent turbines. \par
  Table.\ref{tab2} presents the auxiliary load requirements for a 5 MW alkaline ELZ, obtained from a 240 MW hydrogen production project in Inner Mongolia. Similar to wind farms, not all auxiliary devices operate simultaneously at rated power during black-start. Nevertheless, for worst-case capacity sizing, all potential auxiliary devices are included in the calculation.

\subsection{Transmission losses}
\label{subsec2.2}
\begin{figure}[h]
  \centering
  \includegraphics[width=2.5in]{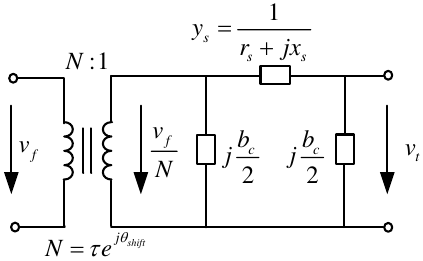}
  \caption{Common $\pi$ equivalent branch model}
  \label{fig3}
\end{figure}
Fig.\ref{fig2} indicates that during black-start, transmission losses primarily originate from transformers, transmission lines, and phase-shifting transformers. As shown in Fig.\ref{fig3}, the system is modeled as an ideal phase-shifting transformer in series with a transmission line, represented using the standard $\pi$-equivalent model \cite{ref28}. The transformer tap ratio is expressed as: 
\begin{equation}
N=\tau e^{j \theta_{\text {shift }}}
\end{equation}
where $\tau$ is the magnitude of the tap ratio and $\theta_{\text {shift }}$ is the phase shift angle. $r_s$, $x_s$ and $b_c$ denote per-unit-length resistance, reactance and susceptance, respectively.\par
The branch admittance matrix is given by:
\begin{equation}
  Y_{b r}=\left[\begin{array}{cc}
  \left(y_s+j \frac{b_c}{2}\right) \frac{1}{\tau^2} & -y_s \frac{1}{\tau e^{-j \theta_{\text {shift }}}} \\
  -y_s \frac{1}{\tau e^{j \theta_{\text {shift }}}} & y_s+j \frac{b_c}{2}
  \end{array}\right]
 \end{equation}\par
This formulation enables accurate power flow analysis during the black-start process, thereby quantifying the active and reactive power required from the PEMFC.
\section{Black-Start Control strategy}
\label{sec3}
\begin{figure}[h]
  \centering
  \includegraphics[width=5.3in]{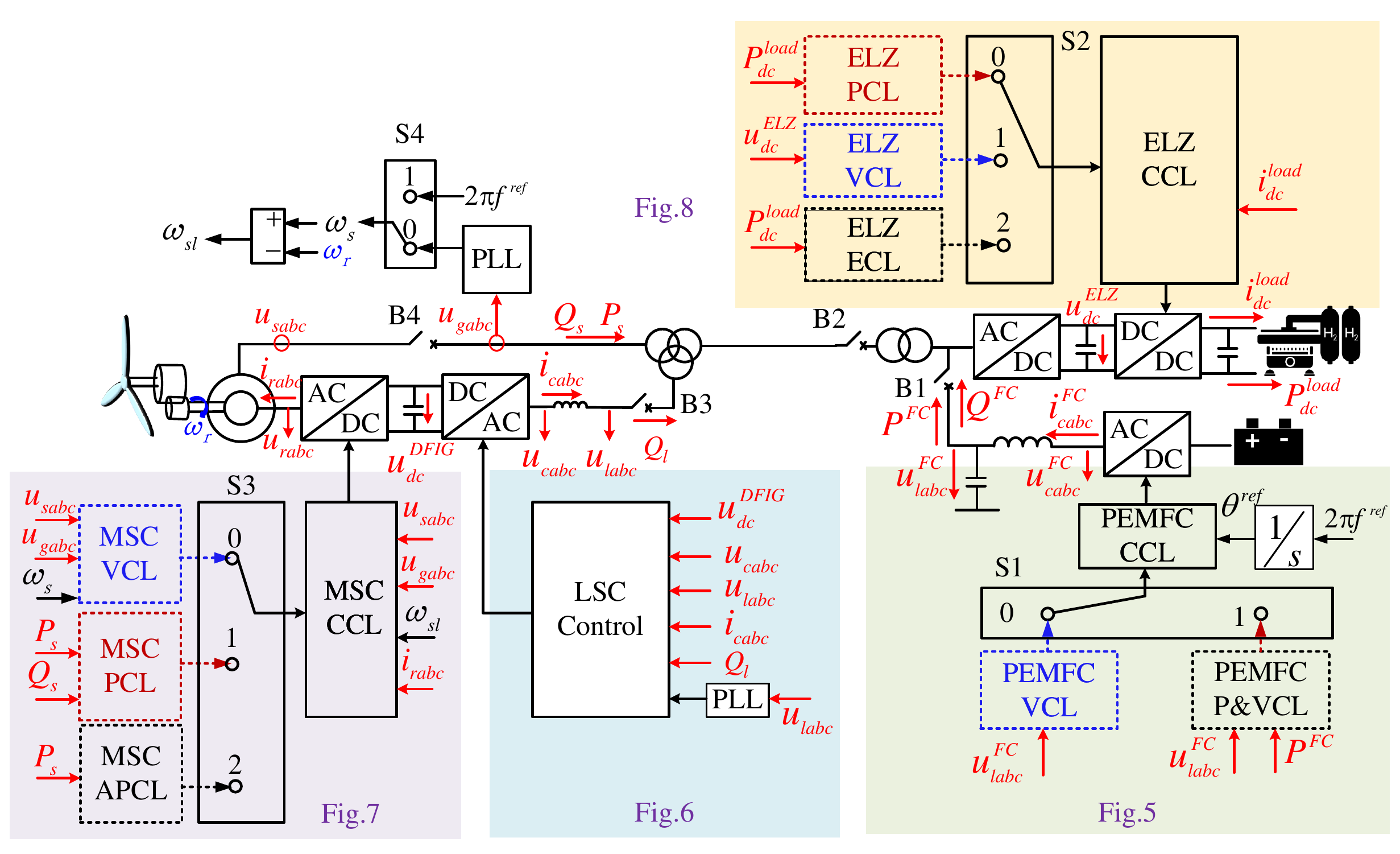}
  \caption{Black-start control diagram of an islanded W2H system}
  \label{fig4}
\end{figure}
There are two existing technical routes for the islanded W2H system: DFIG/PMSG grid-forming and energy storage grid-forming. The drawback of the former is that grid-forming control by wind turbines requires reserving active power, which prevents maximum power point tracking (MPPT) and increase wind curtailment. The latter raises the overall system cost. Both approaches reduce the economic viability of green hydrogen, making islanded W2H systems technically feasible but commercially less attractive. To address these issues, this study proposes two coordinated control strategies: WHCC and HSCC. In WHCC, the hydrogen production unit (HPU) serves as a QV node, balancing active power, while the DFIG acts as a P$\theta$ node, balancing reactive power and maintaining fixed frequency. Notably, this classification introduces novel bus node types not present in conventional power systems. The WHCC control strategy has been discussed in detail in another article by the author \cite{ref27}. In HSCC, the HPU serves as a QV node, while the DIFG operates as a PQ node. The P$\theta$ role is transferred to the PEMFC, which maintains frequency and reactive power balance.

Fig.\ref{fig4} summarizes the block diagrams of these control strategies and their associated black-start control schemes. The following subsections, with reference to Fig.\ref{fig4}, detail the startup procedure of the PEMFC, DFIG and HPU, followed by the complete black-start sequences of the islanded W2H system.
\subsection{Startup Procedure for the PEMFC}
\begin{figure}[h]
  \centering
  \includegraphics[width=4in]{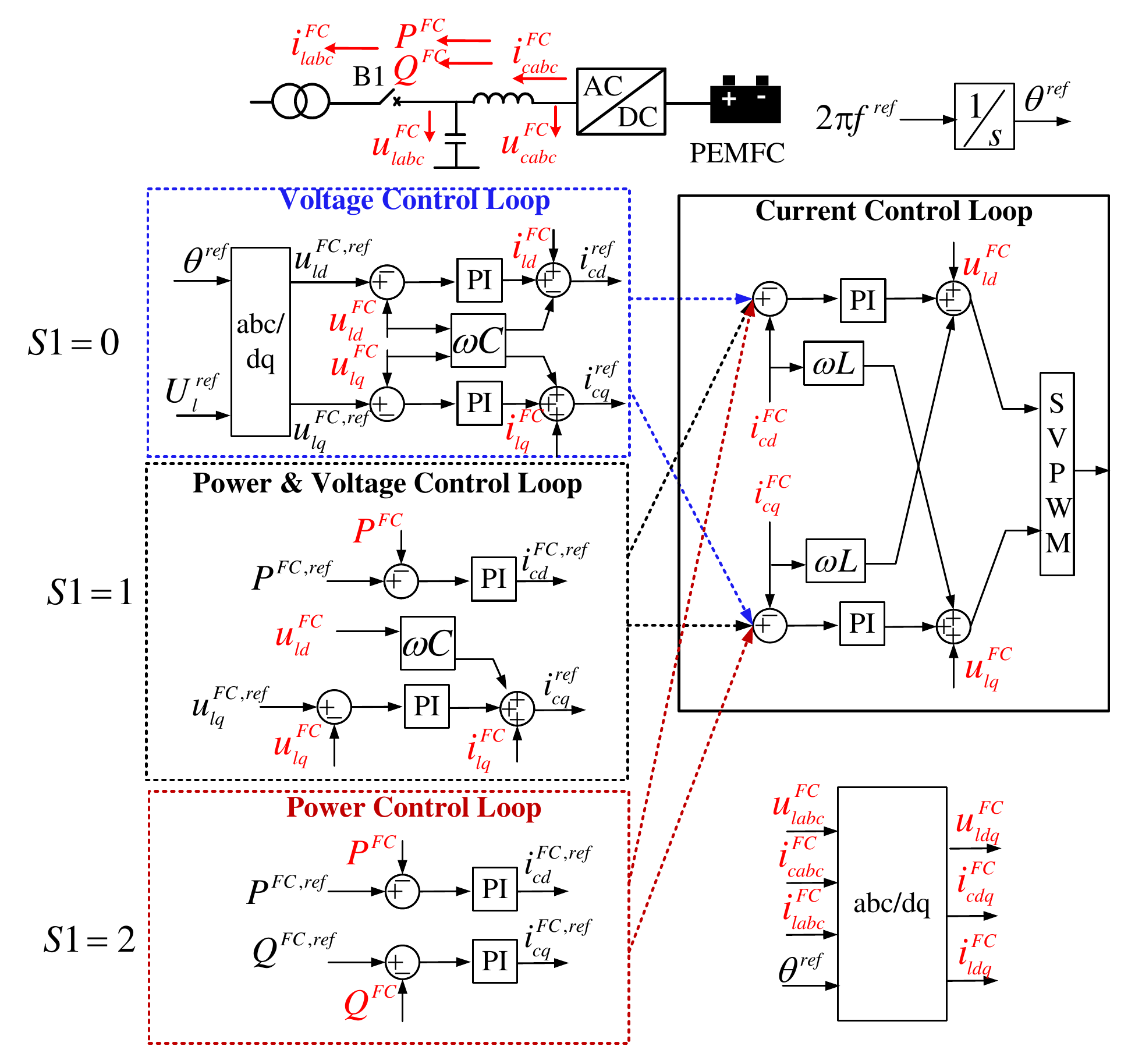}
  \caption{Control diagram of PEMFC}
  \label{fig5}
\end{figure}
Switch $S1$ selects the outer-loop mode of the PEMFC controller. At the initial stage of black-start, $S1 = 0$, activating the voltage control loop (VCL). As shown in Fig.\ref{fig5}, the reference inputs of the VCL are voltage magnitude and frequency, allowing the PEMFC to establish the initial voltage and frequency of the islanded microgrid \cite{ref29}. This stage also charges passive system components and supplies auxiliary power to DFIG1 and HPU1.\par
Once DFIG1 and HPU1 are started, the operating mode of the PEMFC changes. In WHCC, breaker $B1$ is opened, disconnecting the PEMFC. In HSCC, $B1$ remains close and $S1 = 1$, switching the PEMFC controller from VCL to power and voltage control loop (PVCL). In the PVCL and PCL, the reference phase angle $\theta^{r e f}$ still derived from the predefined reference frequency. 

To minimize hydrogen consumption, the PEMFC active power reference $P^{r e f}$ is set to zero. Because the inner loop of PEMFC controler is shared across all modes (VCL, PVCL, and PCL), transitions among these modes remain smooth.
\subsection{Startup Procedure for the DFIG}
\begin{figure}[h]
  \centering
  \includegraphics[width=3.5in]{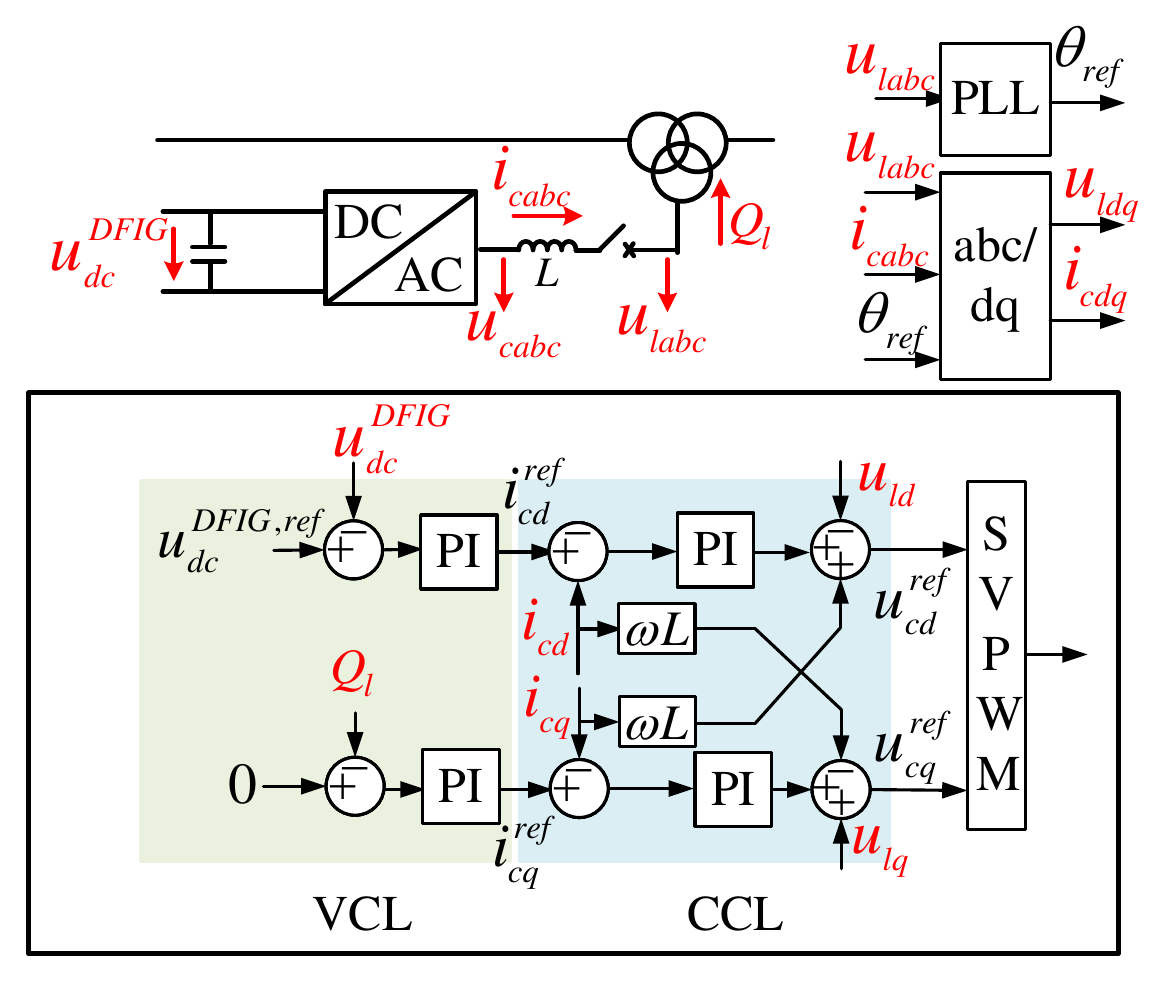}
  \caption{Control diagram of LSC}
  \label{fig6}
\end{figure}
The DFIG employs a back-to-back converter topology, with the machine-side converter (MSC) interfacing with the rotor and the line-side converter (LSC) managing grid interaction. \cite{ref30}. For stability, the LSC must be energized prior to the MSC to ensure DC-link voltage regulation and rotor excitation \cite{ref31}.  \par
\begin{figure}[h]
  \centering
  \includegraphics[width=3.5in]{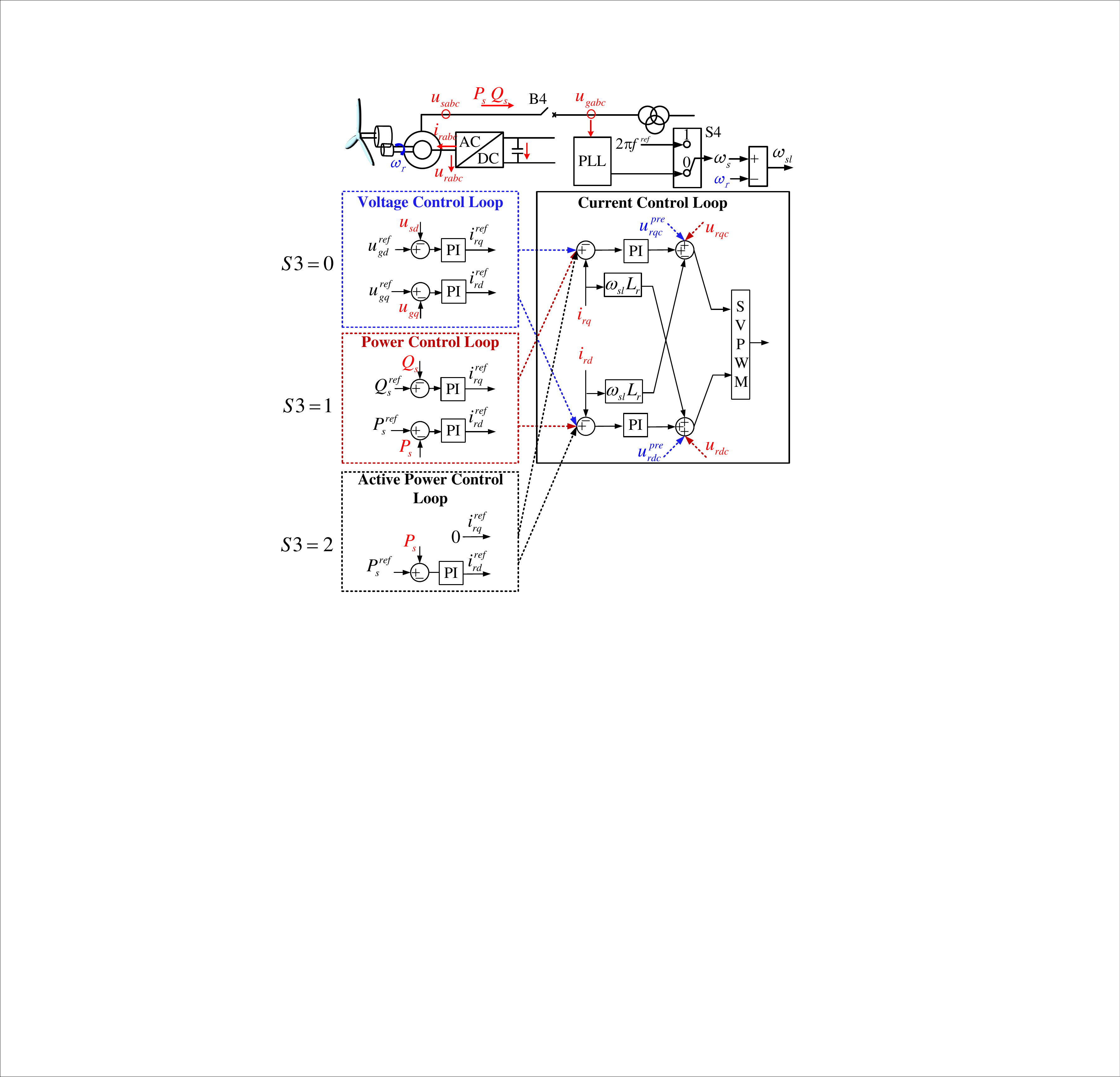}
  \caption{Control diagram of MSC}
  \label{fig7}
\end{figure}

As shown in Fig.\ref{fig6}, the LSC maintains the DC bus voltage at 1150V. Fig.\ref{fig7} illustrates the control diagram of the MSC, at startup, the stator is disconnected from the microgrid.The MSC outer loop operates in VCL, enabling the stator voltage to track the microgrid voltage in magnitude, phase, and frequency, thereby reducing transients during synchronization. After synchronization, the stator connects to the microgrid and the MSC transitions to PCL, regulating DFIG active power. \par
After the HPU completes its startup, switches $S 3$ and $S 4$ shift to different positions depending on the coordinated control strategy. In WHCC, $S 3=2$ and $S 4=1$, the MSC outer control loop regulates active power only, while the DFIG assumes reactive power balancing and frequency regulation as a P$\theta$ node. The MSC controls the stator voltage frequency to be $f^{ref}$. In HSCC, $S 3=1$ and $S 4=0$, the DFIG functions as a PQ node, with frequency and reactive power responsibilities shifted to the PEMFC. It should be noted that the correct terms $u_{r q c}^{p r e}$ and $u_{r q c}$ are the differential values, the detailed calculation procedure can be found in \cite{ref32}.
\subsection{Startup Procedure for the HPU}
\begin{figure}[h]
  \centering
  \includegraphics[width=3.5in]{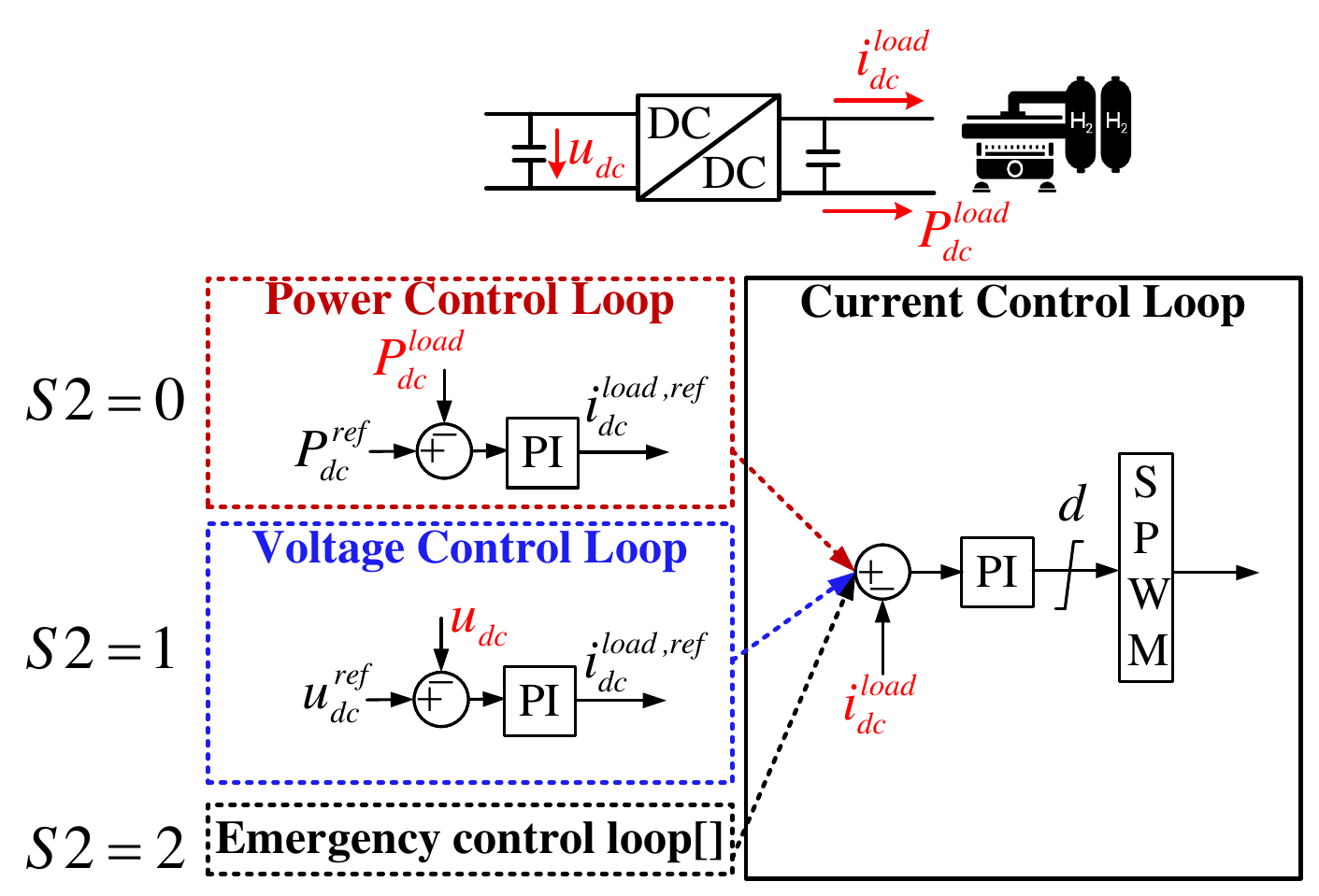}
  \caption{Control diagram of ELZ}
  \label{fig8}
\end{figure}
The HPU employs an uncontrolled rectifier, making its startup relatively simple. Switch S2 determines the outer control loop mode in the ELZ controller. As shown in Fig.\ref{fig8}, initially, $S2=0$, activating the PCL, which ensures the ELZ consumes its minimum operating power. Once the microgrid stabilizes, $S2$ is switched to 1, activating the VCL to regulate the input voltage of the DC/DC converter and stabilize the microgrid voltage. When $S 2$ switches to 2, the ELZ operates in emergency control mode \cite{ref27}.
\subsection{Black Start Sequence of the Islanded W2H System}
Table\ref{tab3} summarizes the black-start sequences under WHCC and HSCC. The first five steps are identical, encompassing PEMFC startup, LSC activation, MSC synchronization, and HPU connection. The key distinction lies in Step 6. In WHCC, the PEMFC is disconnected, and the DFIG assumes the P$\theta$ role for frequency regulation. In HSCC, the PEMFC remains connected, assumes the P$\theta$ role, and maintains frequency and reactive power, while the ELZ operates as an active power slack node.

This structured sequencing ensures reliable black-start, smooth transition to islanded operation, and stable voltage and frequency under fluctuating wind conditions.
\begin{table}[h]
\centering
\caption{Balck Start Sequence of WHCC, HSCC and WHSCC}
\label{tab3}
\renewcommand{\arraystretch}{1.2}
\begin{tabular}{ccccc}
\hline
Step &
  Trigger Conditions &
  \multicolumn{3}{c}{Procedure} \\ \hline
1 &
  \parbox{6.5cm}{Black-start initiation command} &
  \multicolumn{3}{c}{\begin{tabular}[c]{@{}c@{}}S1=S2=S3=S4=0\\ B1=B2=1\\ B3=B4=0\\ PEMFC startup\end{tabular}} \\
2 &
  \begin{tabular}[c]{@{}c@{}}\parbox{6.5cm}{The microgrid voltage and frequency are stabilized}\end{tabular} &
  \multicolumn{3}{c}{\begin{tabular}[c]{@{}c@{}}B3=1\\ LSC startup\end{tabular}} \\
3 &
  \begin{tabular}[c]{@{}c@{}}\parbox{6.5cm}{The system maintains the DC bus voltage at the reference value}\end{tabular} &
  \multicolumn{3}{c}{MSC startup} \\
4 &
  \begin{tabular}[c]{@{}c@{}}\parbox{6.5cm}{The stator voltage is synchronized with the microgrid voltage}\end{tabular} &
  \multicolumn{3}{c}{\begin{tabular}[c]{@{}c@{}}B4=1\\ S3=1\end{tabular}} \\
5 &
  \parbox{6.5cm}{The microgrid reaches steady-state} &
  \multicolumn{3}{c}{HPU startup} \\ \cline{3-4} 
\multirow{2}{*}{\textcolor{red}{6}} &
  \multirow{2}{*}{\parbox{6.5cm}{\textcolor{red}{The microgrid reaches steady-state}}} &
  \textcolor{red}{WHCC} &
  \textcolor{red}{HSCC} &\\ \cline{3-4} 
 &
   &
  \begin{tabular}[c]{@{}c@{}}\textcolor{red}{S2=1}\\ \textcolor{red}{S3=2}\\\textcolor{red}{S4=1}\\ \textcolor{red}{B3=0}\end{tabular} &
  \begin{tabular}[c]{@{}c@{}}\textcolor{red}{S2=1}\\ \textcolor{red}{S1=1}\end{tabular} &
\\ \hline
\end{tabular}
\end{table}
\section{Results and Discussions}
To validate the proposed methodology and control strategies, detailed case studies are conducted in Matlab/Simulink. The black-start power capacity is quantified, and simulations are conducted in MATLAB/Simulink based on black-start sequences presented in Section\ref{sec3}. In the case study, a 62.5 MW wind farm comprising ten 6.25 MW DFIGs is connected to a 50 MW hydrogen production plant, which consists of ten 5 MW electrolyzers. The wind farm and hydrogen plants are 10 km apart from each other and interconnected via a 35 kV overhead transmission line. Detailed parameters of the transformers, transmission line, and other electrical components are listed in \cite{ref33}.
\subsection{Calculation of Black Start Capacity}
\begin{figure}[h]
  \centering
  \includegraphics[width=3.5in]{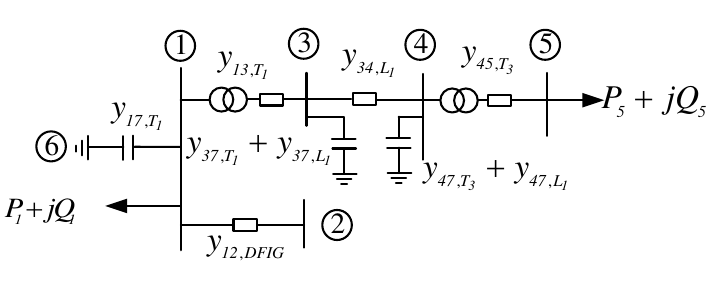}
  \caption{Black-start power supply topology}
  \label{fig9}
\end{figure}
The following assumptions are made in this study:
(1) Because the power supply efficiency can exceed 99\% \cite{ref34}, power losses in the DC/AC, DC/DC, and AC/DC converters are ignored;(2) The transmission line is assumed to be lossless, i.e., the line attenuation factor is zero. Based on these assumptions, the topology of black-start process can be abstracted into the electrical network diagram shown in Fig.\ref{fig9}. Bus 2 represents the DFIG. Branch 3-4 corresponds to the transmission line. A power flow calculation is performed for the six-bus single-machine system. Each transmission branch is labeled with a unique label, and the reference power flow directions are defined such that the positive direction of each branch is from the bus with the lower number to the one with the higher number. $P_1$ denotes the minimum auxiliary load required for wind farm startup, which is assumed to be 5\% of the rated power of a single DFIG unit as mentioned in section \ref{subsec2.1}. $P_5$ and $Q_5$ denote the active and reactive power consumption, respectively, of all auxiliary load within the hydrogen production facility, excluding the HPUs. The branch admittance data and initial bus voltage values are presented in \cite{ref33}.\par
\begin{figure}[h]
  \centering
  \includegraphics[width=4.5in]{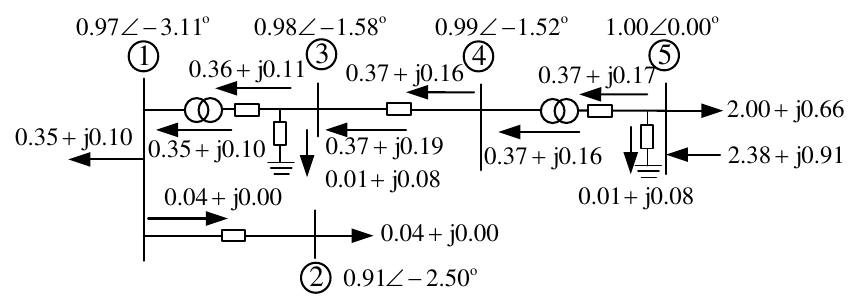}
  \caption{The power flow results}
  \label{fig10}
\end{figure}
Fig.\ref{fig10} shows the results of the power flow analysis, including bus voltages, branch power flows, and system losses. The stator of DFIG is disconnected, and only the LSC connects to the microgrid. Because the LSC is controlled to provide zero reactive power, the bus only requires a small amount of active power to maintain the DC-link voltage. The power flow calculation indicates that approximately 40 kW of active power is needed for this purpose. Bus 5 is defined as the reference bus, as its voltage magnitude and frequency are controlled by the black-start power source via an inverter. Branches 1-3 and 4-5 represent transformer branches. The results show that the no-load reactive power loss due to transformer excitation is approximately 0.08 MVar. Because of   the capacitive effect to ground, branch 3-4 injects approximately 0.03 MVar of capacitive reactive power into the system. In conclusions, the power flow results demonstrate that during the black-start process, the PEMFC and its inverter must be capable of delivering a minimum of 2.38 MW active power and 0.91 MVar reactive power to ensure stability of the system. Considering a 30\% margin, the rated capacity of the PEMFC is determined to be 3 MW.
\subsection{Case study 1: WHCC}
\begin{figure}[h]
  \centering
  \includegraphics[width=4in]{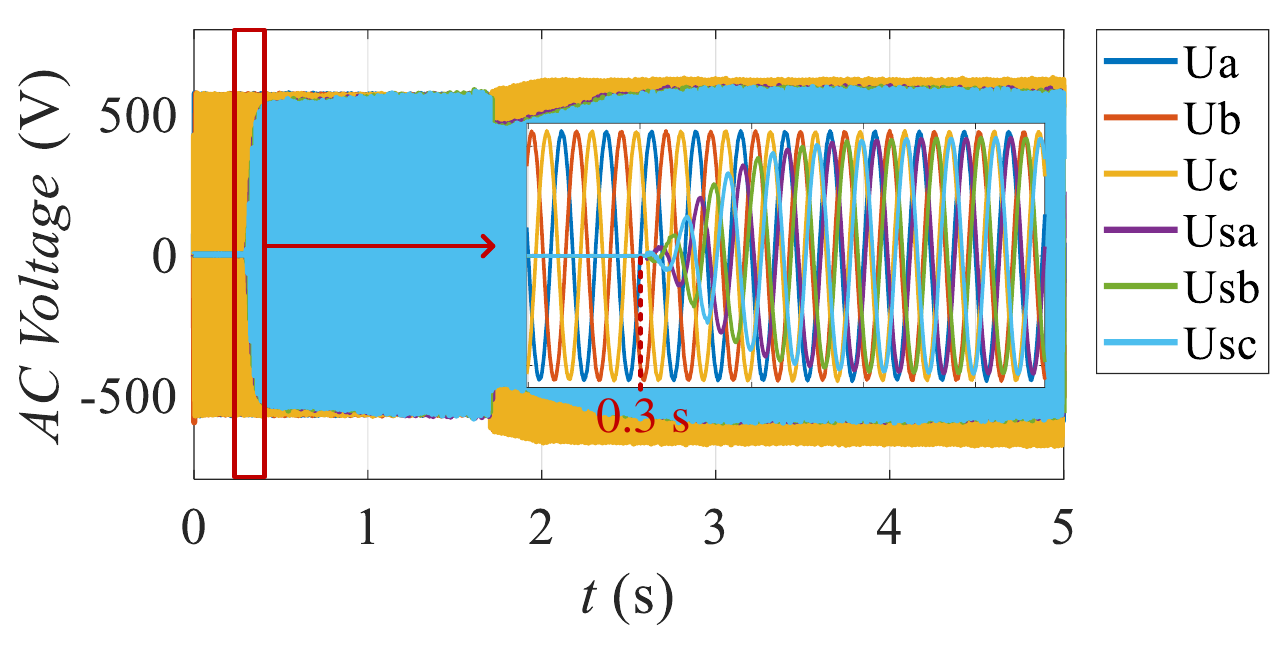}
  \caption{The simulation results of the case study 1: AC voltage at low voltage side}
  \label{fig11}
\end{figure}

\begin{figure}[h]
  \centering
  \includegraphics[width=3in]{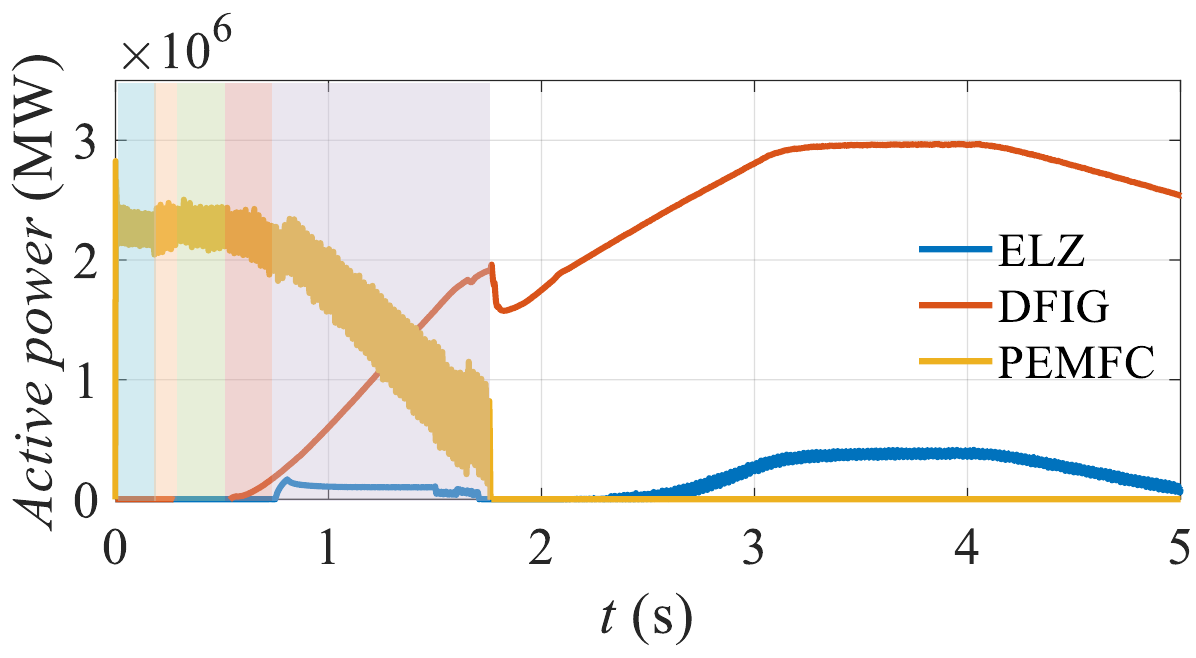}
  \caption{The simulation results of the case study 1: active power}
  \label{fig12}
\end{figure}

\begin{figure}[h]
  \centering
  \includegraphics[width=3in]{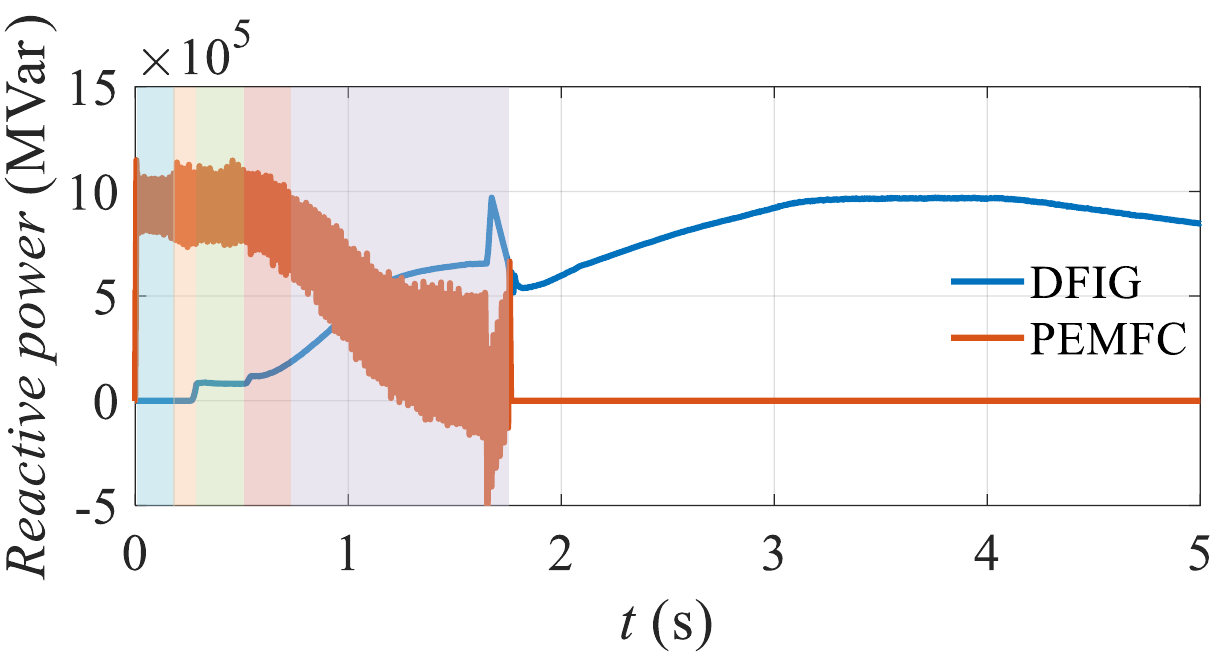}
  \caption{The simulation results of the case study 1: reactive power}
  \label{fig13}
\end{figure}

\begin{figure}[h]
  \centering
  \includegraphics[width=3.5in]{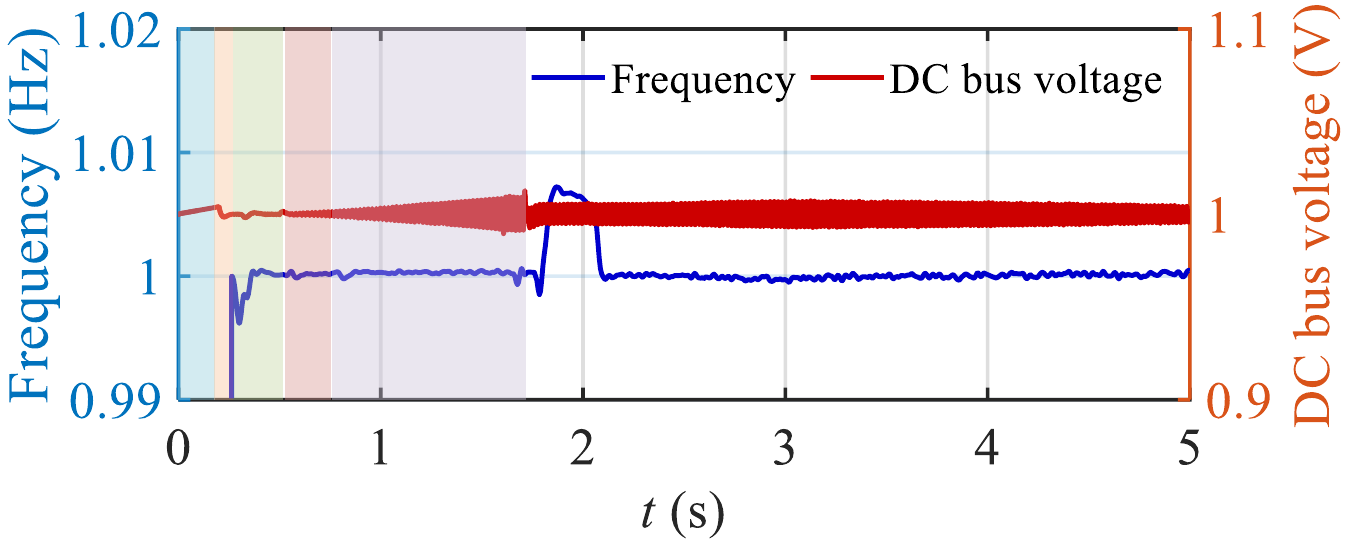}
  \caption{The simulation results of the case study 1: frequency and DC voltage bus}
  \label{fig14}
\end{figure}

Case study 1 simulates the black-start process of the WHCC islanded wind-hydrogen system. The simulation results are presented in Fig.\ref{fig11}-Fig.\ref{fig14}. The distinct background colors correspond sequentially to the black-start sequence presented in Table \ref{tab3}.\par
\textbf{Step 1:} The PEMFC startup sequence initiates before 0.2 seconds, while the outer control loop operates in VLC. The established AC bus voltage demonstrates a frequency of 50 Hz and an amplitude of 690 V, as depicted in Fig.\ref{fig11}. Additionally, the PEMFC concurrently supplies the active power demanded by line losses, DFIG auxiliary systems, and HPU auxiliary systems, along with the reactive power required for transmission line charging, transformer excitation, and DFIG rotor excitation. As shown in Fig.\ref{fig12} and Fig.\ref{fig13},The PEMFC generates active power and reactive power of 2.26 MW and 0.93 Mvar, respectively, exhibiting calculation errors of 5.0\% and 2.2\% relative to the theoretical values presented in Section 4.1.\par
\textbf{Step 2:} When the simulation reaches 0.2 seconds, the LSC startup, establishing the DC bus voltage for the DFIG system. As depicted in Fig.\ref{fig14}, the DC bus voltage reaches a steady-state value of 1150 V after 0.2 seconds.\par
\textbf{Step 3:} When the simulation reaches 0.3 seconds, the MSC initiates its startup sequence. The outer controller operates in VCL, facilitating stator voltage tracking of the AC bus voltage. As shown in Fig.\ref{fig11}, the stator voltage exhibits a gradual increase starting from 0.3 seconds, converging toward the AC bus voltage. As illustrated by the curve with green background in Fig.\ref{fig13}, the DFIG absorbs 0.08 MVar inductive reactive power from the grid during this period to facilitate rotor excitation.\par
\textbf{Step 4:} When the simulation reaches 0.5 seconds, the stator circuit breaker B4 is engaged, integrating the DFIG into the microgrid. The outer controller transitions to PCL, controlling the DFIG's active and reactive power generation based on the predetermined setpoints. Before Step 6, the DFIG active and reactive power setpoints exhibit gradual increments, as illustrated by the curves with red and purple backgrounds in Fig.\ref{fig12} and Fig.\ref{fig13}, respectively. This strategy aims to facilitate DFIG power generation to the system, reducing PEMFC power contribution and mitigating system disturbances during PEMFC disconnection.\par
\textbf{Step 5:} When the simulation reaches 0.7 seconds, the HPU startup, its outer control loop is PCL.\par
\textbf{Step 6:}  When the simulation reaches 1.7 seconds, The PEMFC is disconnected, the HPU outer controller switches to VLC to ensure system voltage stability. The DFIG outer controller transitions to APCL, while the DFIG frequency source switch S4 is engaged to position 1, enabling DFIG-based frequency control. The reference active power is calculated by MPPT algorithm.  As illustrated in Fig.\ref{fig14}, the system frequency exhibits transient oscillations during the frequency source switching process, followed by rapid convergence to a stable operating point. With the black-start sequence completed, the system enters normal islanded operation mode, facilitating the integration of additional DFIGs and HPUs according to wind speed conditions. 
\subsection{Case study 2: HSCC}
\begin{figure}[h]
  \centering
  \includegraphics[width=4in]{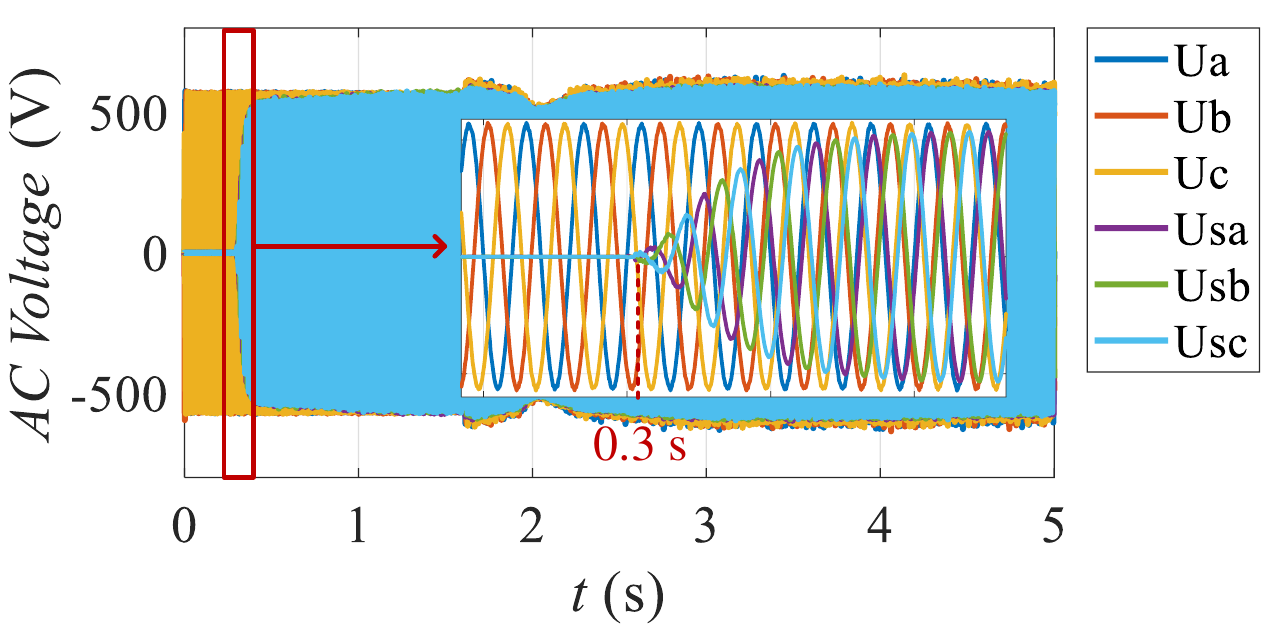}
  \caption{The simulation results of the case study 2: AC voltage at low voltage side}
  \label{fig15}
\end{figure}

\begin{figure}[h]
  \centering
  \includegraphics[width=3in]{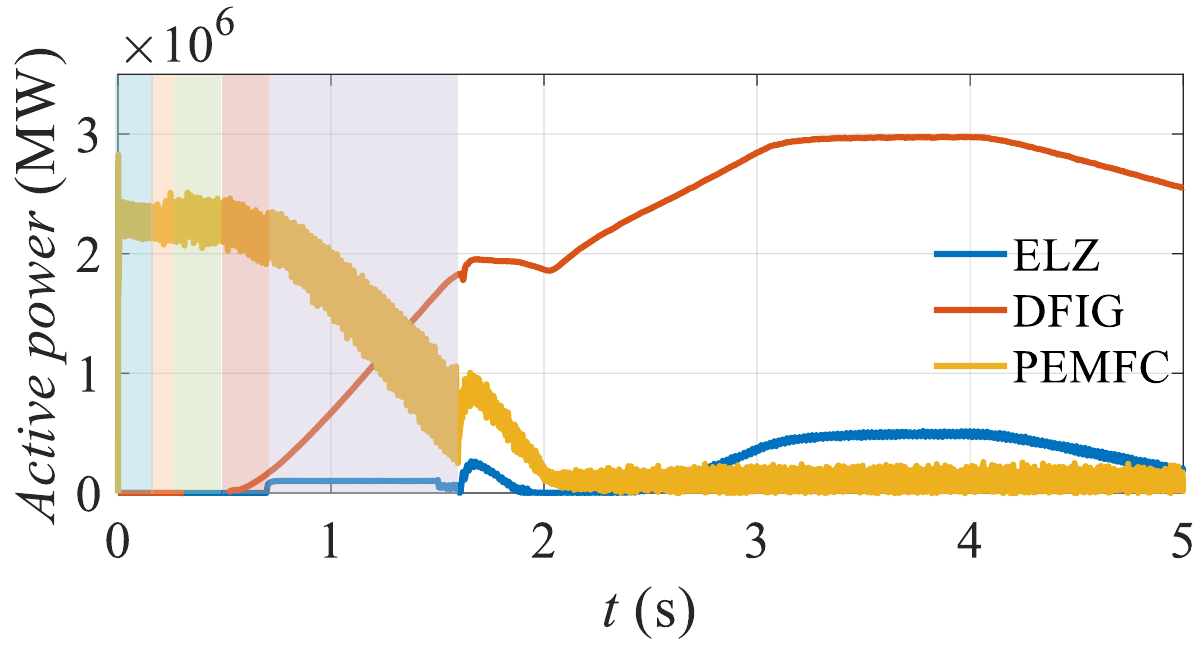}
  \caption{The simulation results of the case study 2: active power}
  \label{fig16}
\end{figure}

\begin{figure}[h]
  \centering
  \includegraphics[width=3in]{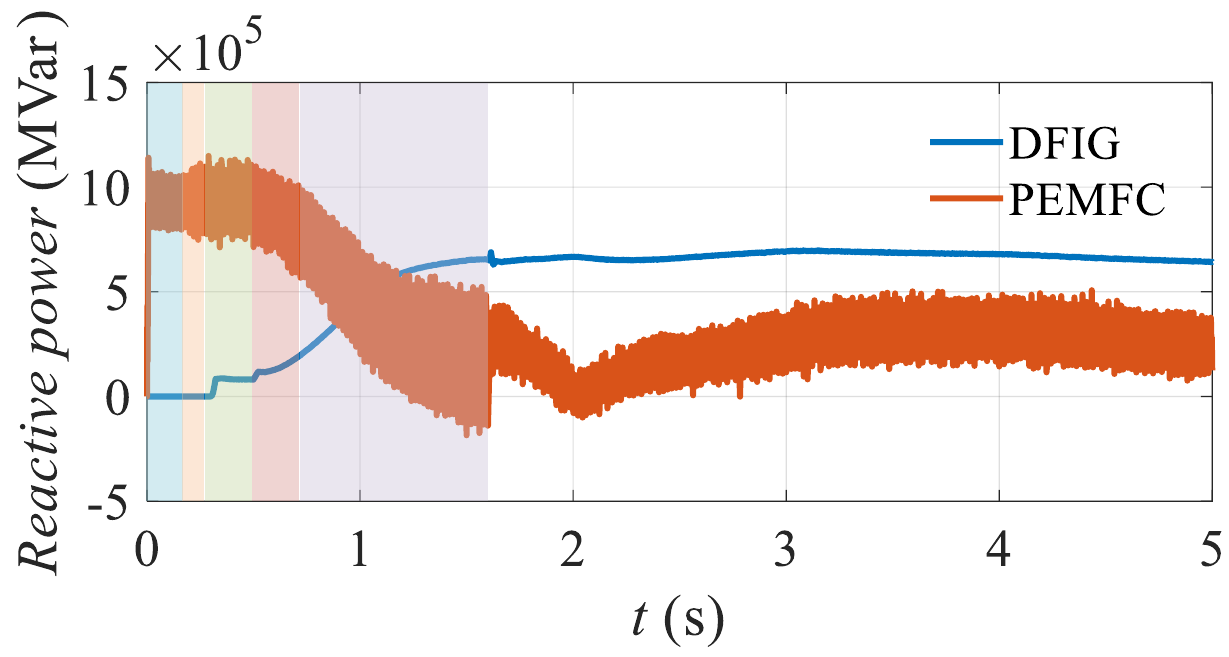}
  \caption{The simulation results of the case study 2: reactive power}
  \label{fig17}
\end{figure}

\begin{figure}[h]
  \centering
  \includegraphics[width=3.5in]{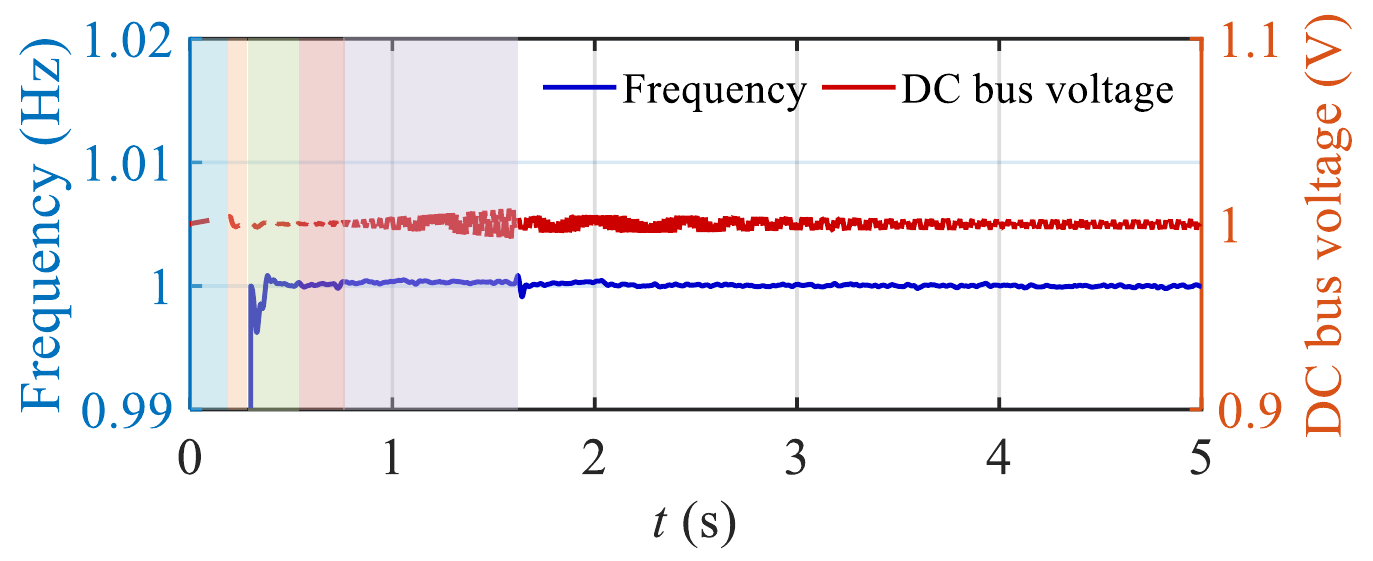}
  \caption{The simulation results of the case study 2: frequency and DC voltage bus}
  \label{fig18}
\end{figure}

Case study 2 simulates the balck start process of the HSCC islanded wind-hydrogen system. The simulation results are presented in Fig.\ref{fig15}-Fig.\ref{fig18}. In the HSCC black-start sequence, Steps 1 to 5 follow the same control strategy as WHCC, with Step 6 being the only distinguishing factor.\par
\textbf{Step 6:}  When the simulation reaches 1.7 seconds, The PEMFC is still connected, but its outer controler switches to P\&VCL. As shown in Fig.\ref{fig16}, the PEMFC generates active power of 50 kW, which is adequate only to compensate for its own switching losses. In contrast to WHCC, the HSCC control strategy employs the PEMFC as a reactive power slack node during this phase, actively compensating for reactive power imbalances between wind turbines and system reactive loads, as represented in Fig.\ref{fig17}. Because the system frequency is consistently maintained by the PEMFC, the HSCC control strategy demonstrates significantly reduced frequency oscillations compared to the WHCC control strategy when transitioning to normal operation mode, as shown in Fig.\ref{fig18}. Concurrently, the ELZ controller switch S2 is engaged to position 1, while the outer control loop switches to VCL. The ELZ functions as an active power slack node during this phase, compensating for system active power imbalances while regulating voltage to rated values, as depicted in Fig.\ref{fig16}. The ELZ active power generation tracks the DFIG active power output variations. While the MSC controller outer control loop maintains its configuration, the active power reference is updated to MPPT power, allowing the DFIG to operate in MPPT mode.
\section{Conclusions}
A systematic methodology for black-start power source capacity calculation
is established. This study proposes HSCC strategy for off-grid W2H systems, and develops comprehensive black-start sequencing protocols for both WHCC and HSCC strategy. Csae studies conducted in MATLAB/Simulink validate the accuracy of the proposed black-start power source capacity sizing methodology,confirm the stability of the HSCC control strategy, and demonstrate the feasibility of both black-start sequences.\par
In the WHCC strategy, after completing black-start and entering normal operation mode, the absence of energy storage equipment reduces control complexity and operational costs. The HSCC strategy provides ELZ manufacturers with a pathway to gain project leadership in renewable hydrogen initiatives, as the key technologies are concentrated on the hydrogen production side, eliminating the need for redevelopment of existing wind turbine control systems. Furthermore, from a system performance perspective, the HSCC control strategy maintains frequency control through PEMFC throughout the entire process, resulting in minimal frequency fluctuations. In contrast, the WHCC system frequency exhibits fluctuations of approximately 1\% during PEMFC connection/disconnection switching in the case study presented in this paper.\par
A significant advantage of the proposed black-start power source capacity calculation approach is its universal applicability, extending from pure off-grid wind-hydrogen systems to process off-grid configurations, with the methodology remaining valid when auxiliary equipment power consumption is excluded from the computational framework.


\appendix
\section{The Parameters of the Case Study}
\label{app1}
\section{The Parameters of Power Flow}
\label{app2}
Appendix text.






\bibliographystyle{elsarticle-num.bst}
\bibliography{ref.bib}

@article{ref1,
  title={Green hydrogen for energy transition: A critical perspective},
  author={Angelico, Ruggero and Giametta, Ferruccio and Bianchi, Biagio and Catalano, Pasquale},
  journal={Energies},
  volume={18},
  number={2},
  pages={404},
  year={2025},
  publisher={MDPI}
}

@techreport{ref2,
  title={Decarbonization in the steel value chain: Global and the Brazilian experiences},
  author={de Paula, Germano Mendes},
  year={2025},
  institution={Working Paper DIP-BR 06/2025, IERI-UFU}
}

@article{ref3,
  title={Redesigning electrification of China’s ammonia and methanol industry to balance decarbonization with power system security},
  author={Li, Jiarong and Lin, Jin and Wang, Jianxiao and Lu, Xi and Nielsen, Chris P and McElroy, Michael B and Song, Yonghua and Song, Jie and Lyu, Xuefeng and Yu, Mingkai and others},
  journal={Nature Energy},
  pages={1--12},
  year={2025},
  publisher={Nature Publishing Group UK London}
}

@article{ref4,
  title={Optimal sizing and pricing of grid-connected renewable power to ammonia systems considering the limited flexibility of ammonia synthesis},
  author={Yu, Zhipeng and Lin, Jin and Liu, Feng and Li, Jiarong and Zhao, Yuxuan and Song, Yonghua and Song, Yanhua and Zhang, Xinzhen},
  journal={IEEE Transactions on Power Systems},
  volume={39},
  number={2},
  pages={3631--3648},
  year={2023},
  publisher={IEEE}
}

@article{ref5,
  title={A study on the promoting role of renewable hydrogen in the transformation of petroleum refining pathways},
  author={Shi, Xiaofei and Wang, Gang and Wang, Xiaolin and Chen, Bo},
  journal={Processes},
  volume={12},
  number={7},
  pages={1317},
  year={2024},
  publisher={MDPI}
}

@article{ref6,
  title={Green hydrogen for energy transition: A critical perspective},
  author={Angelico, Ruggero and Giametta, Ferruccio and Bianchi, Biagio and Catalano, Pasquale},
  journal={Energies},
  volume={18},
  number={2},
  pages={404},
  year={2025},
  publisher={MDPI}
}

@article{ref7,
  title={Optimization of capacity configuration and comprehensive evaluation of a renewable energy electrolysis of water for hydrogen production system},
  author={Zhou, Huairong and Wu, Xin and Li, Chunlei and Yang, Siyu and Chen, Zhichen and Lu, Jun and Fang, Chen},
  journal={Chinese Journal of Chemical Engineering},
  volume={76},
  pages={301--317},
  year={2024},
  publisher={Elsevier}
}

@article{ref8,
  title={On-grid/off-grid operation mode and economic analysis of renewable power to ammonia system},
  author={Lin, J and Yu, Z and Zhang, X and Li, JR},
  journal={Proc. CSEE},
  pages={1--13},
  year={2023}
}

@article{ref9,
  title={Optimal sizing of isolated renewable power systems with ammonia synthesis: Model and solution approach},
  author={Yu, Zhipeng and Lin, Jin and Liu, Feng and Li, Jiarong and Zhao, Yuxuan and Song, Yonghua},
  journal={IEEE Transactions on Power Systems},
  volume={39},
  number={5},
  pages={6372--6385},
  year={2024},
  publisher={IEEE}
}

@article{ref10,
  title={Adapting to limited grid capacity: Perceptions of injustice emerging from grid congestion in the Netherlands},
  author={de Winkel, Eva and Lukszo, Zofia and Neerincx, Mark and Dobbe, Roel},
  journal={Energy Research \& Social Science},
  volume={122},
  pages={103962},
  year={2025},
  publisher={Elsevier}
}

@article{ref11,
  title={Integrated analysis of a hydrogen-based port: Energy, exergy, environmental, and economic sustainability},
  author={Maleki, Farhad and Ledari, Masoumeh Bararzadeh and Fani, Maryam},
  journal={International Journal of Hydrogen Energy},
  volume={100},
  pages={1402--1420},
  year={2025},
  publisher={Elsevier}
}

@article{ref12,
  title={Energy storage sizing by copula modelling joint distribution for wind farm to be black-start source},
  author={Liu, Weipeng and Liu, Yutian},
  journal={IET Renewable Power Generation},
  volume={13},
  number={11},
  pages={1882--1890},
  year={2019},
  publisher={Wiley Online Library}
}

@article{ref13,
  title={Method for the energy storage configuration of wind power plants with energy storage systems used for black-start},
  author={Li, Cuiping and Zhang, Shining and Zhang, Jiaxing and Qi, Jun and Li, Junhui and Guo, Qi and You, Hongfei},
  journal={Energies},
  volume={11},
  number={12},
  pages={3394},
  year={2018},
  publisher={MDPI}
}

@article{ref14,
  title={Grid-forming control strategies for blackstart by offshore wind farms},
  author={Jain, Anubhav and Sakamuri, Jayachandra N and Cutululis, Nicolaos A},
  journal={Wind Energy Science Discussions},
  volume={2020},
  pages={1--22},
  year={2020},
  publisher={G{\"o}ttingen, Germany}
}

@article{ref15,
  title={Blackstart from HVDC-connected offshore wind: Hard versus soft energization},
  author={Jain, Anubhav and Sabor{\'\i}o-Romano, Oscar and Sakamuri, Jayachandra N and Cutululis, Nicolaos A},
  journal={IET renewable power generation},
  volume={15},
  number={1},
  pages={127--138},
  year={2021},
  publisher={Wiley Online Library}
}

@article{ref16,
  title={Investigation of black-starting and islanding capabilities of a battery energy storage system supplying a microgrid consisting of wind turbines, impedance-and motor-loads},
  author={Marchgraber, J{\"u}rgen and Gawlik, Wolfgang},
  journal={Energies},
  volume={13},
  number={19},
  pages={5170},
  year={2020},
  publisher={MDPI}
}

@article{ref17,
  title={Grid-forming VSM control for black-start applications with experimental PHiL validation},
  author={Alassi, Abdulrahman and Feng, Zhiwang and Ahmed, Khaled and Syed, Mazheruddin and Egea-Alvarez, Agusti and Foote, Colin},
  journal={International Journal of Electrical Power \& Energy Systems},
  volume={151},
  pages={109119},
  year={2023},
  publisher={Elsevier}
}

@article{ref18,
  title={Performance evaluation of a bess unit for black start and seamless islanding operation},
  author={Izadkhast, Seyedmahdi and Cossent, Rafael and Fr{\'\i}as, Pablo and Garc{\'\i}a-Gonz{\'a}lez, Pablo and Rodr{\'\i}guez-Calvo, Andrea},
  journal={Energies},
  volume={15},
  number={5},
  pages={1736},
  year={2022},
  publisher={MDPI}
}

@ARTICLE{ref19,
  author={Liu, Weipeng and Liu, Yutian and Wu, Lei},
  journal={IEEE Transactions on Sustainable Energy}, 
  title={Model Predictive Control Based Voltage Regulation Strategy Using Wind Farm as Black-Start Source}, 
  year={2023},
  volume={14},
  number={2},
  pages={1122-1134},
  doi={10.1109/TSTE.2023.3238523}}

@article{ref20,
  title={Novel black start control strategy with power boundary analysis for PMSG-based wind turbines},
  author={Huang, Guohang and Shen, Feifan and Huang, Sheng and Wu, Qiuwei and Wei, Juan and Qu, Yinpeng and Wang, Pengda},
  journal={IEEE Transactions on Sustainable Energy},
  volume={15},
  number={2},
  pages={1224--1238},
  year={2023},
  publisher={IEEE}
}

@article{ref21,
  title={A model predictive control based generator start-up optimization strategy for restoration with microgrids as black-start resources},
  author={Zhao, Yuxuan and Lin, Zhenzhi and Ding, Yi and Liu, Yilu and Sun, Lei and Yan, Yong},
  journal={IEEE Transactions on Power Systems},
  volume={33},
  number={6},
  pages={7189--7203},
  year={2018},
  publisher={IEEE}
}

@article{ref22,
  title={Integrating black start capabilities into offshore wind farms by grid-forming batteries},
  author={Pagnani, Daniela and Kocewiak, {\L}ukasz and Hjerrild, Jesper and Blaabjerg, Frede and Bak, Claus Leth},
  journal={IET Renewable Power Generation},
  volume={17},
  number={14},
  pages={3523--3535},
  year={2023},
  publisher={Wiley Online Library}
}

@article{ref23,
  title={Black-Start Strategy for Offshore Wind Power Delivery System Based on Series-Connected DRU-MMC Hybrid Converter},
  author={Li, Feng and Chen, Danqing and Chen, Honglin and Luo, Shuxin and Yu, Hao and Hou, Tian and Wang, Guoteng and Huang, Ying},
  journal={Electronics},
  volume={14},
  number={13},
  pages={2543},
  year={2025},
  publisher={MDPI}
}

@article{ref24,
  title={Coordinated Control of Grid-Forming Wind Turbines and Grid-Forming Energy Storage Systems for Power System Restoration},
  author={Zhang, Yuping and Xie, Yunyun and Cai, Sheng and Wu, Qiuwei and Zhu, Haobin and Xiang, Zhengrong},
  journal={IEEE Transactions on Sustainable Energy},
  year={2025},
  publisher={IEEE}
}

@article{ref25,
  title={Decentralized frequency control for black start of full-converter wind turbines},
  author={Asensio, Andres Pena and G{\'o}mez, Santiago Arnaltes and Rodriguez-Amenedo, Jose Luis and Cardiel-{\'A}lvarez, Miguel {\'A}ngel},
  journal={IEEE Transactions on Energy Conversion},
  volume={36},
  number={1},
  pages={480--487},
  year={2020},
  publisher={IEEE}
}

@article{ref26,
  title={Black-start capability of DFIG wind turbines through a grid-forming control based on the rotor flux orientation},
  author={Rodr{\'\i}guez-Amenedo, Jos{\'e} Luis and G{\'o}mez, Santiago Arnaltes and Mart{\'\i}nez, Jes{\'u}s Castro and Alonso-Martinez, Jaime},
  journal={IEEE Access},
  volume={9},
  pages={142910--142924},
  year={2021},
  publisher={IEEE}
}

@article{ref27,
  title={Achieving Stability and Optimality: Control Strategy for a Wind Turbine Supplying an Electrolyzer in the Islanded Storage-less Microgrid},
  author={Yang, Bosen and Ma, Kang and Lin, Jin and Zhang, Mingjun and Ji, Zhendong and Liu, Zhi and Song, Yonghua and others},
  journal={arXiv preprint arXiv:2501.08853},
  year={2025}
}

@book{ref28,
  title={Power system dynamics: stability and control},
  author={Machowski, Jan and Lubosny, Zbigniew and Bialek, Janusz W and Bumby, James R},
  year={2020},
  publisher={John Wiley \& Sons}
}

@article{ref29,
  title={Control interaction modeling and analysis of grid-forming battery energy storage system for offshore wind power plant},
  author={Zhao, Fangzhou and Wang, Xiongfei and Zhou, Zichao and Harnefors, Lennart and Svensson, Jan R and Kocewiak, {\L}ukasz Hubert and Gryning, Mikkel Peter Sidoroff},
  journal={IEEE Transactions on Power Systems},
  volume={37},
  number={1},
  pages={497--507},
  year={2021},
  publisher={IEEE}
}

@article{ref30,
  title={Reactive power capability of wind turbines based on doubly fed induction generators},
  author={Engelhardt, Stephan and Erlich, Istvan and Feltes, Christian and Kretschmann, J{\"o}rg and Shewarega, Fekadu},
  journal={IEEE Transactions on energy conversion},
  volume={26},
  number={1},
  pages={364--372},
  year={2010},
  publisher={IEEE}
}

@article{ref31,
  title={DFIG topologies for DC networks: A review on control and design features},
  author={Marques, Gil D and Iacchetti, Matteo Felice},
  journal={IEEE Transactions on Power Electronics},
  volume={34},
  number={2},
  pages={1299--1316},
  year={2018},
  publisher={IEEE}
}

@book{ref32,
  title={Doubly fed induction machine: modeling and control for wind energy generation},
  author={Abad, Gonzalo and Lopez, Jesus and Rodriguez, Miguel and Marroyo, Luis and Iwanski, Grzegorz},
  year={2011},
  publisher={John Wiley \& Sons}
}

@electronic{ref33,
    url     =   {{https://github.com/ybs22/BS_APPENDIX}},
    title   =   {{Apendix}}
}

@article{ref34,
  title={A state-of-the-art review on soft-switching techniques for DC--DC, DC--AC, AC--DC, and AC--AC power converters},
  author={Mohammed, Sadeq Ali Qasem and Jung, Jin-Woo},
  journal={IEEE Transactions on Industrial Informatics},
  volume={17},
  number={10},
  pages={6569--6582},
  year={2021},
  publisher={IEEE}
}
\end{document}